# Pressure-Induced Critical Influences on Workpiece-Tool Thermal Interaction in High Speed Dry Machining of Titanium

H.A. Abdel-Aal[1] and M. El Mansori

*Laboratoire de Me´canique et Proce´de´s de Fabrication (LMPF), ENSAM CER Cha^lons-en-Champagne, Rue Saint Dominique BP 508, 51006 Chalons-en-Champagne, France*

*[1]corresponding author, hisham.abdel-aal@ensam.fr*

**Abstract.** Cutting tools are subject to extreme thermal and mechanical loads during operation. The state of loading is intensified in dry cutting environment especially when cutting the so called hard-to-cut-materials. Although, the effect of mechanical loads on tool failure have been extensively studied, detailed studies on the effect of thermal loads on the deterioration of the cutting tool are rather scarce. In this paper we study failure of coated carbide tools due to thermal loading. The study emphasizes the role assumed by the thermo-physical properties of the tool material in enhancing or preventing mass attrition of the cutting elements within the tool. It is shown that within a comprehensive view of the nature of conduction in the tool zone, thermal conduction is not solely affected by temperature. Rather it is a function of the so called thermodynamic forces. These are the stress, the strain, strain rate, rate of temperature rise, and the temperature gradient. Although that within such consideration description of thermal conduction is non-linear, it is beneficial to employ such a form because it facilitates a full mechanistic understanding of thermal activation of tool wear.

## INTRODUCTION

Titanium is the fourth most abundant structural metal in the earth's crust. It forms the base for many alloys with remarkable mechanical properties that renders it attractive for many engineering designs. Titanium is known to be a difficult-to-cut due to a combination of strength at elevated temperatures and poor thermal dissipation. This combination leads to the formation of a small Heat Affected Zone (HAZ) while machining [1], and to a short contact length between chip and tool. Consequently, high temperatures dominate material removal of titanium and its alloys. Elevated temperatures are a major catalysis of many tool wear mechanisms [2,3]. Further, they promote adhesion between tool and workpiece and between chips and tool whence leading to the formation of build-up layers. The detrimental effects of temperature elevation in material removal highlight the importance of detailed studies of thermal interaction in the active material volume. Traditionally, studies of such a topic have been confined to analytical [3-5], numerical [6-8], and experimental [9-16] determination of the temperature rise during machining. Critical consideration of the literature, however, reveals that despite it's crucial influence on the mechanics of material removal, physics of thermal energy transport within the critical areas of contact (tool-workpiece and tool-chip) is by far still unexplored. This is despite experimental evidence that infer the destructive effect that inefficient heat removal, from the active material volume, may have on tool wear and quality of the resulting surface [17-22]. Recent results [23, 24] indicate that dissipation of thermal loads acting on the tool-workpiece influences both tool life and the mechanics of machining. Further, it has been shown [25, 26] that the effective values of thermal transport properties for the tool-workpiece combination assume a major role in thermal activation of tool wear. Consequently, understanding thermal catalysis of tool wear should be based on the investigation of: a) factors that induce a change in thermal and electrical transport properties during operation, and b) the effect of that change on the energy kinetics within the material volume directly in contact with the tool.

Studying the change in thermal conductivity is important on two counts. Firstly, the active material volume is the incubator for the potentially generated chips and the potentially generated surface. Baric and thermal histories, within this zone, are likely to affect: the mechanics of chip creation and surface generation. Secondly, elevated stress state (pressure) is characteristic of the zone located directly under the tool. Materials under high pressure, exhibit different behavior than that exhibited under atmospheric pressure. Elevated pressure also induces changes in the thermal and electrical properties materials. For metals, the thermal and the electrical conductivity are related through the well Known Weidman -Franz- Lorenz law (WFL). The consequence of such a relationship on, wear mechanisms is important on the count that pressure induced changes in thermal transport properties will induce a reciprocal change in electrical transport properties. Thus if the pressure induced effect is to reduce thermal conductivity, the reciprocal effect that follows from the WFL is to increase electrical resistivity. This may be very well related to certain types of wear, and to adhesion, in addition to material transfer (the so called build-up-layer) from the work piece to the tool rake side (all of which are observed in machining of titanium and titanium based alloys). Such a connection, to the Knowledge of the authors, is not addressed within machining or tribology literature.

This paper introduces a numerical study of thermal response of titanium in the tool-work piece active zone. The object of investigation is the behavior of the thermal conductivity in the zone of the workpiece that is directly located under the tool. This is accomplished through simulating a cutting cycle, (i.e., the time interval of generating a chip and a new surface). The cycle starts when the tool initially engages the workpiece and terminates when the chip is generated and a new surface is created. Through each time step, we trace the evolution of pressure and temperature fields in the zone located directly under the tool. Subsequently, we feed this data into experimentally obtained maps of electrical and thermal conductivities at different pressures and temperatures. In this fashion, a complete map of the thermal and electrical conduction fields under the tool is obtained for each time step. These maps allow relating experimentally observed wear mechanisms to the mode of energy transport likely to take place based on the evolution of transport property variation.

## VARIATION OF THERMAL CONDUCTIVITY IN SOLIDS

The physics of material behavior at high pressure is established in the most part (at least for metals). There is also an abundance of information and accumulation of comprehensive works (notably the seminal work of Bridgeman [00] and that of Bundy and coworkers [00]) that detail behavior of solids under elevated pressures. However, none of such data have found its way to engineering models (except perhaps in case of semi-conductor processing where pressure induced phase transformations play a key role). Even though, in that context, high pressure considerations are confined to phase identifications and not to determine the different properties at the new phases [99]. This data infer that at an elevated pressure transport properties are significantly different from those at standard conditions. As such in a machining process where elevated pressures and temperatures dominate the active volume of the work-piece one should expect multi-property zones in which local baric and temporal variations may lead to complex responses. At any rate, given the extreme contrast between the value of atmospheric pressure and the actual pressure in the material volume located under the tool one should not expect the active zone of the material to behave like the bulk. However, the decision of whether or not to consider elevated pressure effects on material response hinges on the magnitude of change in the properties of interest at the dominant high pressure.

### General Considerations

Factors affecting thermal conductivity of solids fall within two distinct parameter groups: a thermodynamic group (such as temperatures, pressures, strains, temperature elevation etc., or the so called thermodynamic forces); and an “extrinsic” parameter group that includes influences such as impurities, defects, or bounding surfaces, grain sizes etc.,. Thermal conductivity, and thereby electrical conductivity, will exhibit changes in response to changes in any of one group constituent parameters individually or collectively. In metals, the thermal conductivity changes in response to variations in many factors. Most familiar is the temperature-induced variations. Depending on the metal, and the alloying elements if present, temperature variations may cause the thermal conductivity to decrease, increase, or show mixed behavior in particular temperature intervals. For example, Iron (Fe), Copper (Cu), and their alloys, exhibit a drop in the conductivity with temperature elevation. On the other hand, Nickel (Ni), and Ni-based alloys, including stainless steels, exhibit an increase in their thermal conductivity in response to increase in temperature [2]. Effective strain rate also affects thermal and electrical conductivity [26-28]. This, according to Radinov and Goncharev [ 131], is due to the connection between strain rate and the rate of defect creation in the solid. A change in the rate of defect creation may cause changes in electrical resistively as well [329-12].

# Pressure Induced Influences

Pressure influences both thermal and the electrical conductivity of metals [ ]. Such an effect is never investigated in the context of machining (and tribology in general) despite its perceived crucial influence on the thermal response of both tool and workpiece. Figure 1 (a-through d) depicts the variation in the thermal conductivity and electrical resistivity as a function of temperature and pressure for Commercially Pure Titanium CPT. Thermal conductivity data were extracted from electrical resistivity measurements obtained by Balog and Secco [ ], using the WFL. The figure indicates that at standard conditions, atmospheric pressure and room temperature, the thermal conductivity is at a maximum value. The conductivity drops with pressure increase. The decrease in thermal conductivity is around 35% from the value at atmospheric pressure. It is also noted that the conductivity partially recovers (increases), past a pressure of approximately 2 GPa. At all pressures, the thermal conductivity exhibits an infliction point in variation with temperature. The critical temperature at which infliction takes place is not constant at all pressures. It displays slight variation, around one hundred degrees (see figure 1-c), with pressure.

Variation with temperature is similar for all pressures. At room temperature, the conductivity is at a maximum, then it decreases with temperature increase. There is a critical temperature, associated with the α-β transition [?], beyond which the conductivity starts to increase with temperature elevation. That is, the conductivity of CPT displays an infliction point (or a local minimum) at the critical temperature. The increase in pressure, at room temperature, causes the initial value of the conductivity at the particular pressure contour to drop. Observe, for instance, the variation of the conductivity with pressure for the curve labeled $T_{room}$ in figure 1-d. Note that while the conductivity at room temperature, $K_{o,}$ is approximately 22 W/mC, it drops to $15 < K < 17$ W/mC in the pressure range $0.2<P< 3$ GPA.

Machining entails the application of high pressures (stresses) to remove material and thereby generate surfaces. The bulk of that pressure generally affects the material volume directly located under the cutting tool. This causes the contact Zone between the tool and the work piece to undergo an elevated state of stress. The stresses within this zone may reach 2-4 GPa in case of titanium. Under such a pressure, the affected material volume, which is also the incubator of the potentially generated surface, will exhibit different transport properties than those of the bulk. Despite the existence of comprehensive data that characterizes behavior at elevated pressures for many alloys, incorporation of such an effect is yet to find its way in engineering models (especially in machining) [31-34]. The reason for such a situation is perhaps two fold. Firstly, variation of thermal properties with pressure is a less familiar occurrence to many engineers. Secondly, it is due to the false notion that since nearly all manufacturing processes take place within atmospheric conditions, local pressure variation will not induce significant changes on properties of solids. Whence, there is no need to accommodate pressure effects on thermal behavior when studying material removal processes.

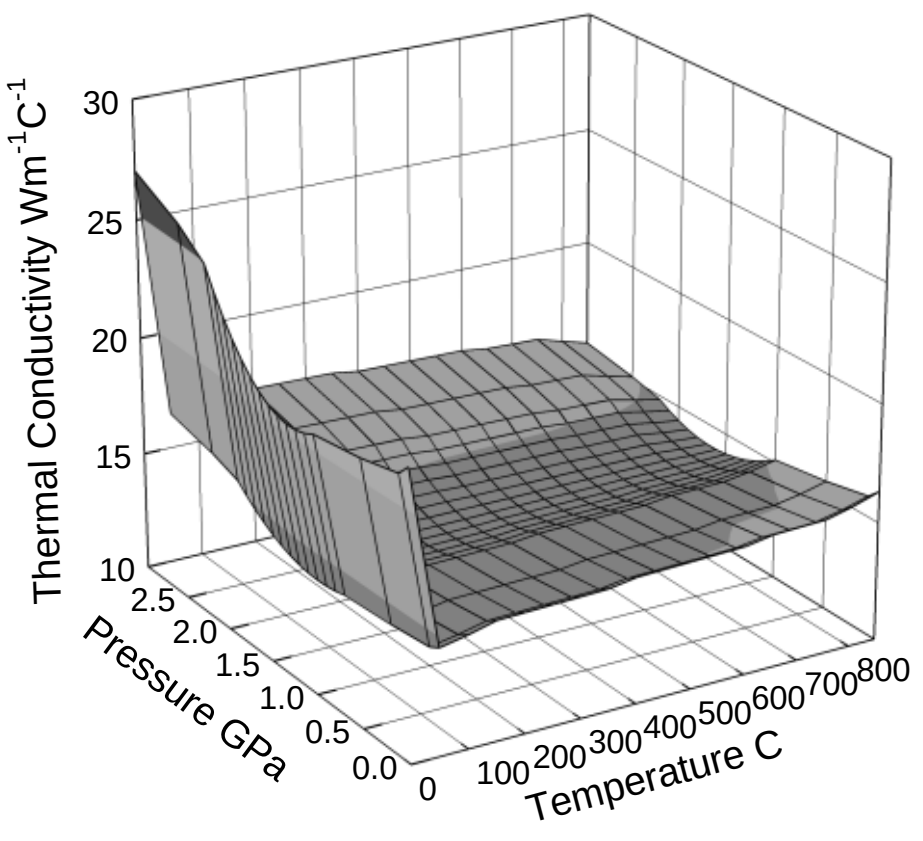

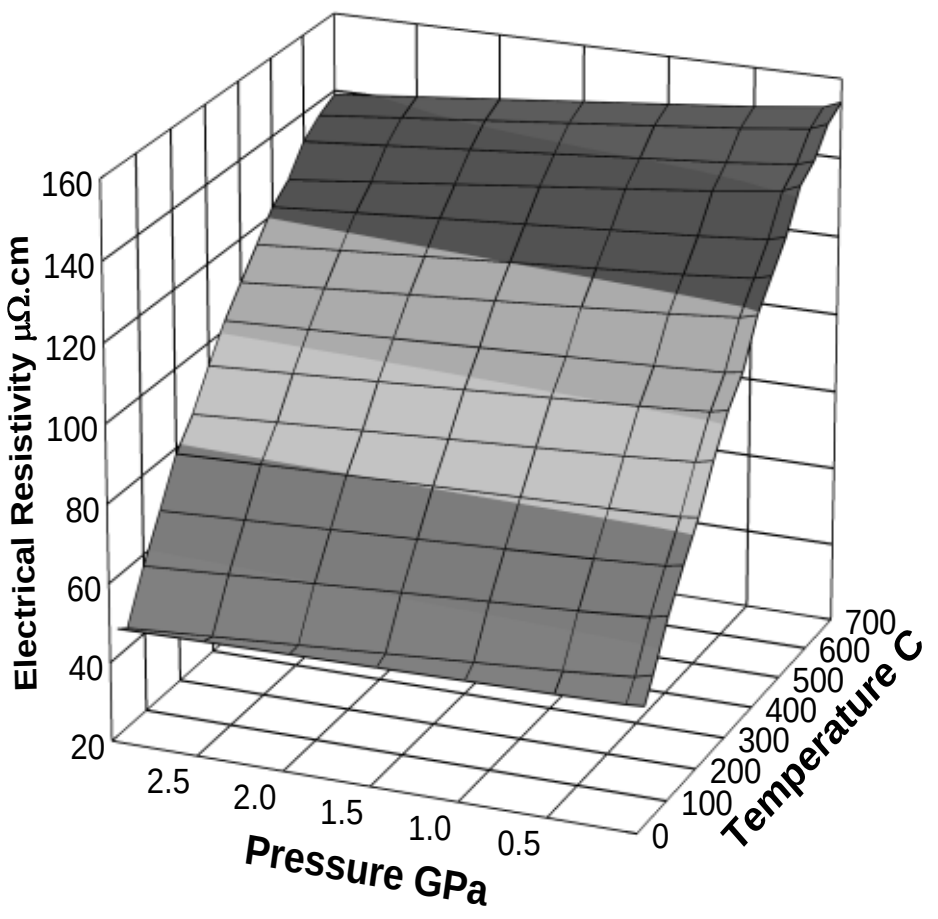

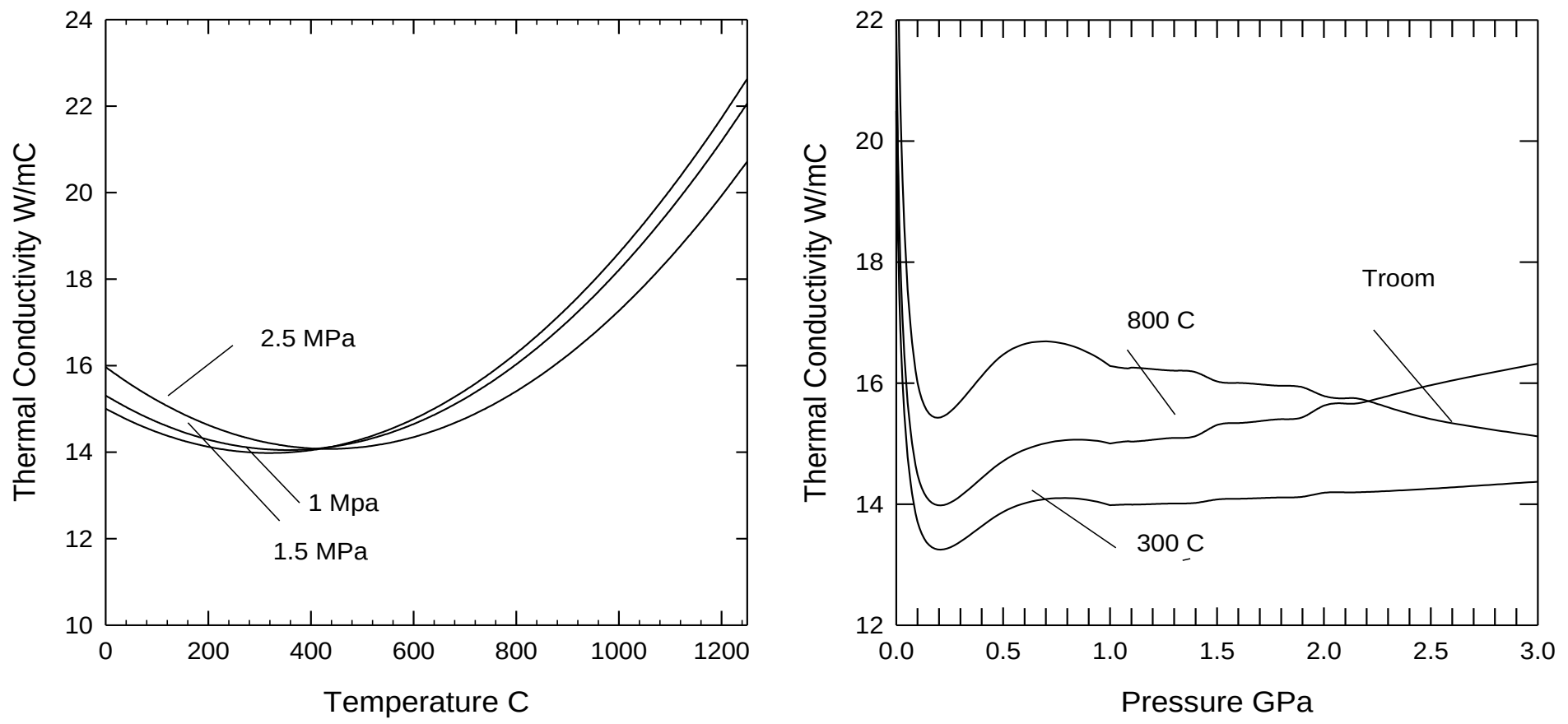

**FIGURE 1**: Variation in the thermal and electrical transport properties of commercially pure Titanium at different pressures and temperatures

However, comparing the value of atmospheric pressure (0.1 MPa) to the pressures reached under the tool in machining titanium (and hard-to-cut materials in general), 2-3 GPa, emphasizes how misleading is such a notion.

## MODELING OF THE THERMAL CONDUCTIVITY

In machining, the tool and the workpiece, experience pressure, strain and temperature rise of considerable rate and magnitude. In addition, the synergetic effects of the cutting forces, temperatures, and mechanical properties result in a coupling between the mechanical and the thermal states of the tool and the workpiece. This coupling evolves during operation due to evolution of the contact parameters. These, in turn, change in response to changes in contact topography. Changes in topography, meanwhile, modify the contact parameters which consequently change the thermo-physical and electrical properties of the contact. As such the active zone in both tool and workpiece will continue to evolve. Under such conditions, the materials involved will exhibit thermo-mechanical properties that are consistent with their evolved state not with those of the original nominal state (i.e., before contact). Measuring the new properties is rather challenging, and in most cases if successful, will entertain uncertainty that are, at least, of the same order of magnitude of computed properties. To this effect modeling is attractive.

Modeling of the thermo-physical properties under actual contact conditions allows the consideration of individual measurements (performed when only one factor induces variation e.g., temperature or pressure-on thermal conductivity) into a working formula that predicts behavior when most of the parameters induce variations. For example, in this work, we consider the effective value of the thermal conductivity of Commercially Pure Titanium (CPT). Knowing the variation of the conductivity due to temperature change and that due to pressure change, we can use statistical methods to develop a working formula that predicts the effective thermal conductivity when pressure and temperature vary simultaneously.

In the most general case we define an effective thermal conductivity $K_{eff}$ given by:

$$K_{eff} = K\left(\sigma, \eta, e^{o}, T, T^{o}\right) \qquad (1)$$

The apparent thermal conductivity is a parameter that incorporates the changes in the thermal conductivity due to the different influences and the coupling between them. Further, it reflects the quality of thermal conduction under actual loading conditions. Depending on available experimental data and the level of sophistication desired in the analysis one may choose the form of the function $K_{eff}$ ($\eta$, $e^{o}$, $T^{o}$, T), in equation one to be linear or a of higher order.

### Modeling Changes Induced by Thermodynamic Forces

To address changes of the thermal conductivity induced by the thermodynamic parameter group, we assume that the variation is a superposition of two contributions thermal and thermo-mechanical. The first represents the variation due to the temperature rise whereas the second represents the variation due to the thermodynamic forces (excluding pressure (or alternatively stress), and the thermo-mechanical coupling. Detailed mathematical and physical justifications for these choices are considered beyond the scope of this presentation; however they may be referred to elsewhere [43-46]. Accordingly, the thermal conductivity is represented by,

$$K_{app} = K(T) - \frac{C_p \, \eta \, e^o}{\alpha \, (\nabla^2 T)} \qquad (2)$$

Where: K(T) the temperature dependant contribution, $C_p$ is the heat capacity, $e^o$ is the strain rate and η is the so called thermo-mechanical coupling factor given by:

$$\eta = \frac{3 K_b \, \alpha^2 \, T}{C_p} \qquad (3)$$

Where: $K_b$ is the bulk modulus of the material. Expressing the temperature effect on the conductivity in the form, $K(T) = K_o \ (1 + \beta \ T)$, where β is the temperature coefficient of the conductivity; replacing the term $\nabla^2 T$, as a first approximation, by T° (rate of temperature rise); and substituting for η from equation (2), the apparent thermal conductivity may be rewritten as:

$$K_{app} = K_o \left[ \ 1 + \Phi T \ \right] \qquad (4)$$

Here Φ is a modified coefficient of conductivity that reflects the combined effects on the point wise variation in the modes of loading and is given by:

$$\Phi = \beta - \frac{3 \alpha K_b D e^o}{K_o T^o} \qquad (5)$$

Equation (4) is a working formula that allows the estimation of the apparent thermal conductivity within the various zones of the tool, given that the other parameters involved are characterized. The main implication of equation 4 however is that the quality of thermal conduction in the tool active zone is affected by the degree of coupling between the thermal and the mechanical states of the tool-workpiece materials in addition to the physical transport properties of the materials involved. Note also that the parameters $\alpha$ and $\dot{e}$ are in essence tensors and can be used to study the multidirectional variation of conduction efficiency.

## Modeling Changes Induced by Stress

If the available data are those pertaining to the variations induced by pressures and temperature equation one may be written as:

$$K_{eff} = K(\sigma, T) \qquad (6)$$

It is understood, however, that pressure and stress could be used interchangeably in equation two.

To evaluate the shape of the shape of the function K (P,T) in equation two, data of thermal conductivity introduced in figure one was subject to statistical analysis. Analysis was performed using a commercially available software package to fit the data to a parabolic form. Such a shape was chosen due to the proximity of the original curves, shown in figure 1a, to the chosen form. The analysis resulted in a working equation that describes variation in conductivity due to simultaneous thermal and pressure effects in the form:

$$K_{eff} = Y + aP + bT + cP^2 + dT^2 \qquad (7)$$

Where Y is a constant and a,b, c and d are regression coefficients. Table 1 provides the magnitudes of the regression coefficients along with the relevant statistical parameters.

**Table 1**: Values of the regression coefficients for calculation of thermal conductivity variation under the combined influence of pressure and temperature elevation

| Parameter | Value | Std. Err | CV(%) |
|---|---|---|---|
| y | 4.69E+01 | 3.51E+01 | 7.48E+01 |
| a | -3.30E-03 | 2.34E-03 | 7.08E+01 |
| b | 3.25E+00 | 3.50E+00 | 1.08E+02 |
| c | 6.16E-06 | 3.72E-06 | 6.04E+01 |
| d | 8.07E-02 | 8.65E-02 | 1.07E+02 |

Equation (3) combined to values of pressures and temperatures that are typical of machining should enable complete mapping of conduction and thermal loading fields in operation. In this work values of pressure and temperature are obtained from finite element simulation. The details of the calculations are detailed elsewhere [99]. In what follows, however, we provide details that are most pertinent to this work.

## Finite Element Modeling

A commercial finite element code *'Third Wave Advantage ®"* was used for simulations. The package uses Coulomb's law for friction and a power strain-hardening law to describe the material behavior. In the calculations the flow stress of the work piece is assumed to be affected by strain rate. Material properties also allow for thermal softening.

Simulations were performed for a K-type cemented carbide insert with three cutting edges. The calculations allow for the presence of a 3 μm thick coating of TiAlN. The tool geometry was first drawn then imported into the code. Table 2 presents a summary of the properties used to describe the insert. The work piece was chosen as Commercially Pure Titanium (CPT), the properties of which were supplied by the FE code.

The software uses a 2D Lagrangian explicit finite element analysis. Simulations utilized a triangle element with a-three quadrature point to mesh the tool-chip structure. A sample mesh is given in figure 3. During simulations, the material is deformed in the primary shear zone. This is assumed to be a thin layer that is inclined to the free surface of the workpiece at the shear angle. The material moves from the primary to the secondary shear zone. This causes dissipation of energy in friction at the contact zones (tool-chip contact at the rake face and tool-workpiece contact at the flank face), and generates frictional heat which further affects the tool surface (crater wear and flank wear). Also during simulations, the mesh, which becomes distorted around the cutting edge is updated by refining both large and small elements. For all simulation runs the total cutting length (input parameter) is kept constant at 2 mm; with this value steady state cutting (for cutting force and temperature levels) is obtained, and a reduction of computing time is ensured.

Based on previous experimental experience, the cutting conditions presented in table 3 were selected for numerical simulation. For all calculations the depth of cut was kept constant to 1.5 mm.

**Table 2**. Material's properties for K-type carbide insert used in simulations

| Material | % weight max | C. L. T ($10^{-6}$/k) | Density Kg/cm$^3$ | T (°c) | ($H_v$) | E N/mm$^2$ |
|---|---|---|---|---|---|---|
| WC | 96% | 5.1 | 15.6 | 2900 | 2150 | $700\times10^3$ |
| Co | 6% | 12.3 | 8.9 | 1495 | - | $100\text{-}180\times10^3$ |
| TiAlN | 3μm | 9.2 | 5.43 | 2950 | 2000 | $260\times10^3$ |

**Table 3**. Cutting conditions used in numerical simulations.

| Cutting Speed $V_c$ (m/min) | 15 | 30 | 60 | 100 |
|---|---|---|---|---|
| Feed rate $f$ (mm/rev) | 0.1 | 0.2 | 0.25 | 0.2 |
| Depth of cut $a_p$ (mm) | 1.5 | 1.5 | 1.5 | 1.5 |

## RESULTS

In what follows we present those results of the simulations that pertain to the subject of this work. The order of the presentation is the following. We start by presenting general features of chip formation and mechanics of surface generation. These are accompanied by an exposition of the evolution of the contact parameters, pressure, temperature and rate of heat generation. Subsequently, we present extracted data for three points that envelope the tool radius, one is at the rake side, the second is under the tool nose, and the third is at the flank side. For each of these points we present the evolution of pressure and temperature within the chip generation cycle. The presented pressure and temperature data are then used to evaluate the thermal conductivity at the as a function of time within the contact cycle. These data in turn is subsequently used to deduce the electrical resistivity fields under the tool.

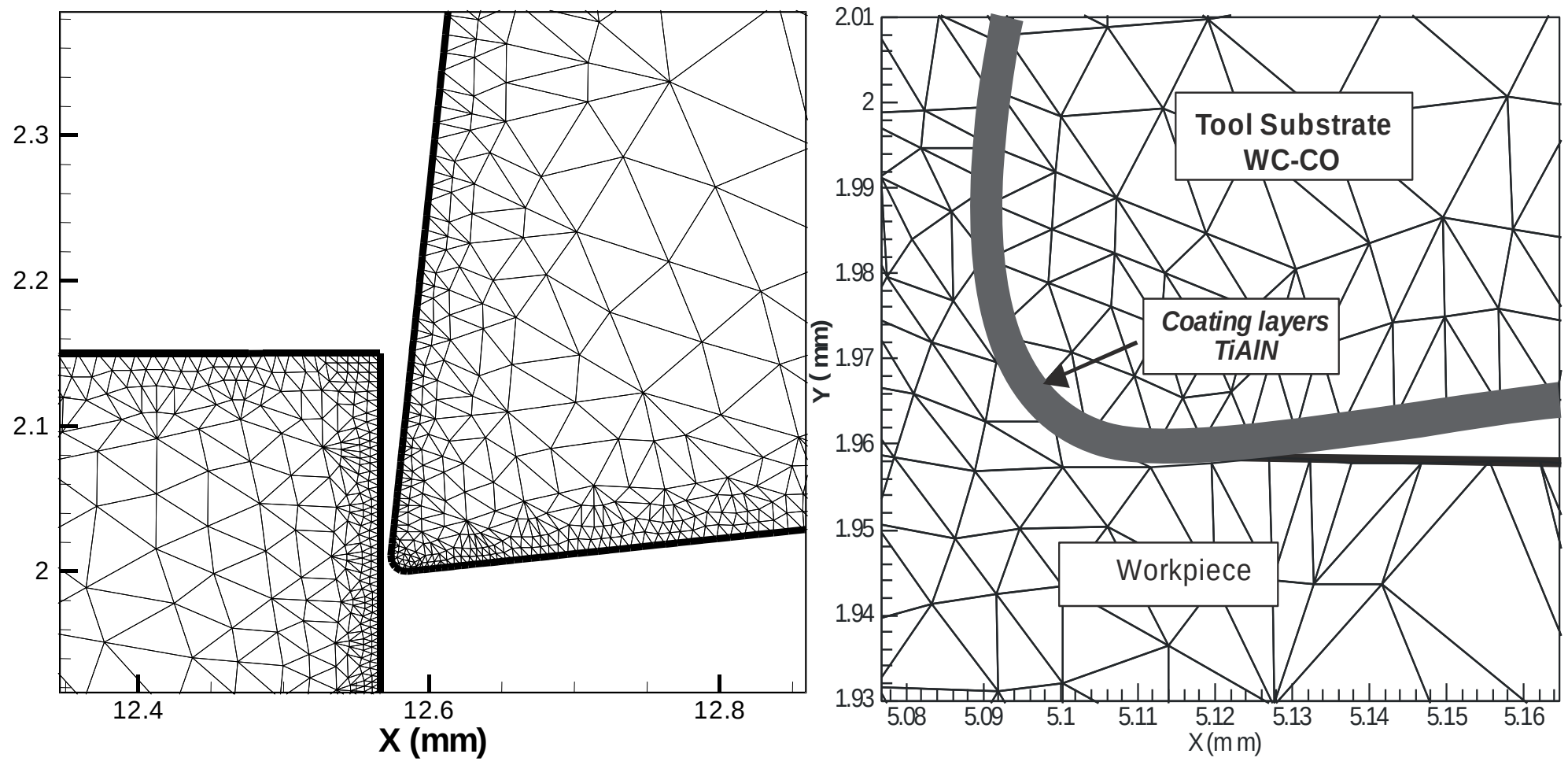

**FIGURE 3**. Meshed tool and workpiece. Tool with 3µm coating (material 1 (substrate): tungsten carbide (*WC-Co*), material 2: (*TiAlN)*.

## Chip Generation Cycle

Figure 4 presents the results of simulation of the chip generation process. The figure contains nine frames labelled (a through i). Each of these frames represents a time interval of about 2.4 E-04 seconds. This particular simulation run was performed for a speed of cut Vc equal to 100 m/min and a feed of 0.2 mm/tooth, with a tool that has 15 degrees rake angle. The frames are magnified so as to focus on the chip-tool interface. Initially we may note that the chip undergoes a “birth cycle”. The first phase of that cycle is termed here the *conception phase.* This is represented by the first three frames (a through c) of figure 4. This phase is where the outline of the chip is formed and perhaps the character of the chip-tool interaction is outlined. The second phase termed here as the *incubation phase*, represented by the following three frames, d through f, of the figure. This phase is where the chip starts to slide and both adhesion and material transfer between workpiece and tool takes place. This phase also is conceived to be where craters on the tool may form. The third phase that compliments the cycle is termed here as the release phase. In such a phase the final detachment of the chip from the workpiece takes place. Here it is envisioned that the energetics of the workpiece-tool interaction may or may not promote residual effects within the work piece (e.g, residual stresses, phase transformations etc.,). It is to be emphasized that cycle comprising these three phases is repeated. However, each cycle begins with the initial conditions, for the particular cycle, that are affected by the preceding history of the contact.

The discussion that follows herein will make use of figures 4 through 7. These represent the chip generation frames, the evolution of: the temperatures, figure 5, the pressure under the tool from the workpiece side, figure 6, and the evolution of the rate of heat release figure 7. Frames in all these figures are in one-to-one correspondence with the chip generation frames.

## Conception Phase

The shear plane is not fully developed. Initial approach of the tool is accompanied by steep temperature elevation. There is a high concentration of hot zones under the tool nose from the workpiece side. A very hot zone (temperature in the order of 1000 C) nucleates under the tool within the workpiece. Compressive pressure is high as well, in excess of 2Gpa under the tool approximately within the hot zone. The workpiece under the tool experiences steep pressure gradients. At this initial stage and due to the pressure elevation, the workpiece material will shift behaviour from that which is familiar under standard conditions to a response that is consistent the pressure elevation that it experiences. It is of interest to note that at this initial stage, the length of contact between chip and tool is to be determined. Such a length will be a function of thermal and electrical environment dominant. These in turn are functions of the thermal and electrical properties within this zone. It is conceived that the intense heat release combined to the pressure induced change in thermal and electrical properties, an environment that is conducive to sticktion and adhesion may form. Note also the relative relief in temperature intensity once the overall form of the generated chip begins to emerge.

Despite that the temperature concentration under the tool is partially relieved, the overall magnitude of the temperature remains high (around 900 C). Relief at this stage originates from the motion of chip on the rake face. This initiates the so called material convection action due to a fresh portion of the workpiece meeting the tool nose. As a result temperature is relieved. However, due to that relative motion heat also is generated. Consistent with the provisions of the theory of moving sources of heat [55] material convection causes some heat to move in a direction opposite to that of sliding. This is reflected in the shift of high temperature contours toward the flank face. Such an action may be related to residual thermal stresses in the workpiece. Under the tool we note that there exists uniform temperature distribution that may extend into the primary shear zone.

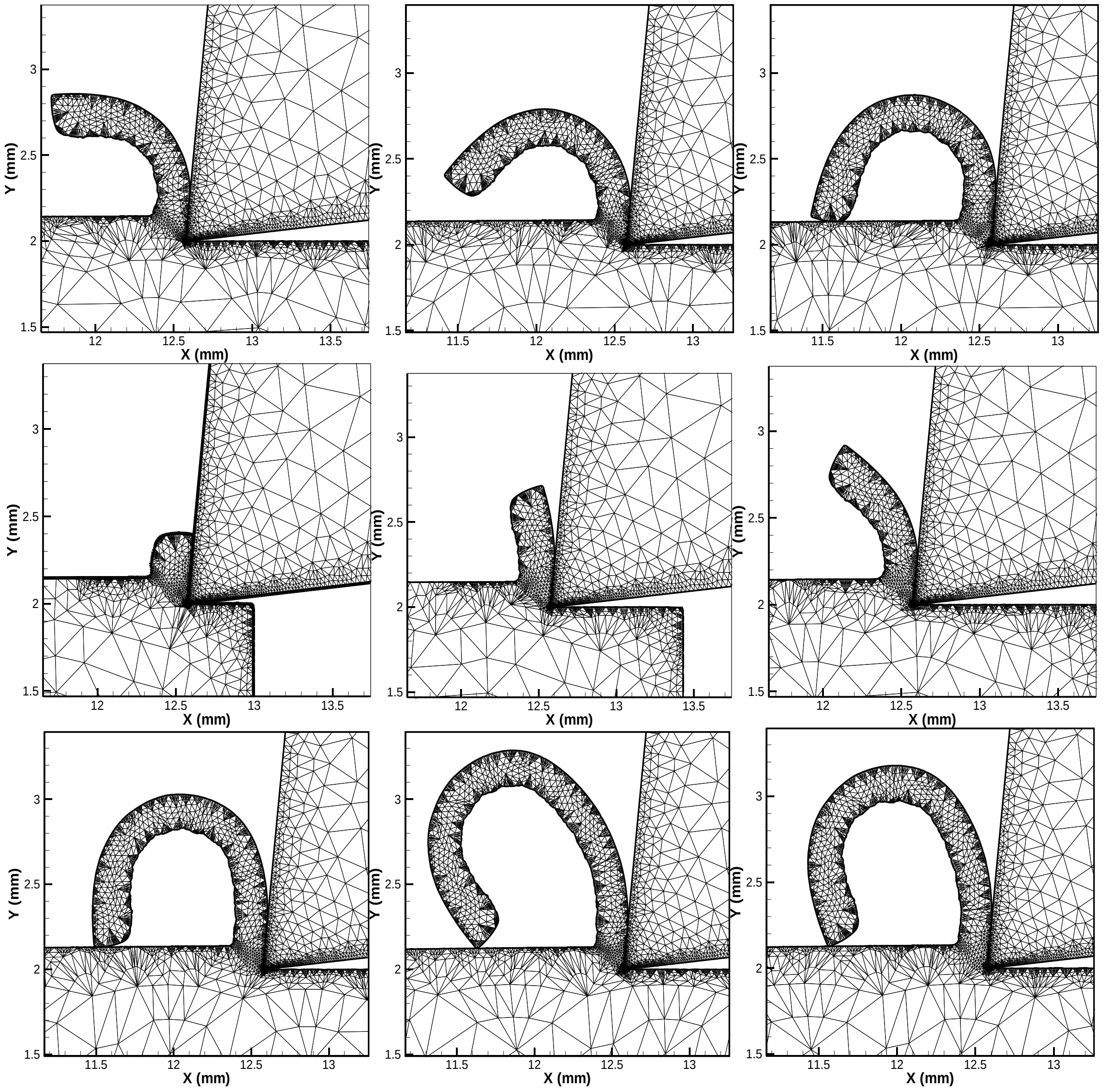

## Incubation Phase

This phase is characterized by dominant material convection effects. This leads to uniform temperature distribution around a rim-shaped zone in the immediate vicinity of the tool nose contour. The magnitude of the temperatures, however, remains high (above 900 C). Close to the rake face, meanwhile, high temperatures in the order of,1000 C, develop. This temperature seems to be localized within the same spot for the duration of the chip creation cycle. It is conceived that at such a high temperature adhesion and chemical reactions in rates that may cause crater wear will be active. Moreover, at such temperature electrical resistivity will be high. This again may promote tribo-electrification mechanisms and contribute to material transfer through adhesion.

The pressure within this phase gets higher, perhaps due to relative material softening. Of interest, however, is that pockets of very high pressure start to nucleate at critical locations within the workpiece active zones. The nucleation initially starts within a minute distance away from the tool contour (perhaps a manifestation of local plasticity), then migrates, on the tool workpiece interface toward the rake face. The magnitude of pressure within these pockets is the highest within the active zone of workpiece material (2 Gpa $\leq$ P $\leq$ 2.5 GPa).

Heat release continues due to material relative motion and compression in a relatively less intense manner. Never the less, one can notice that the intense heat release activity migrates toward the flank face of the tool (particularly where the workpiece is in direct contact with the tool). Again pockets of intense heat release may be identified at the root of the primary shear zone from the flank face side.

## Chip Release Phase

In this phase there is a slight drop in the temperature. However, the contour of the workpiece, closer to the tool nose radius, seems to entrain isothermal conditions where the temperature remains in the neighbourhood of 900 C. The temperature contour of the highest magnitude (in excess of 1000 C) continues to reside within the general area of the rake face root. The pressure meanwhile, continues to drop slightly although high pressure pockets, which seem to occupy bigger areas, continue to nucleate within the material located in the immediate vicinity of tool nose. Again, the magnitude of the pressure within these pockets is around the 2.5 GPa threshold. Heat release activity surges and the zones of higher hear release intensity migrate from the root of the flank face to the general area directly under the tool nose. These quasi-stagnant pockets display an intense rate of heat release (compared to that exhibited during the conception phase.

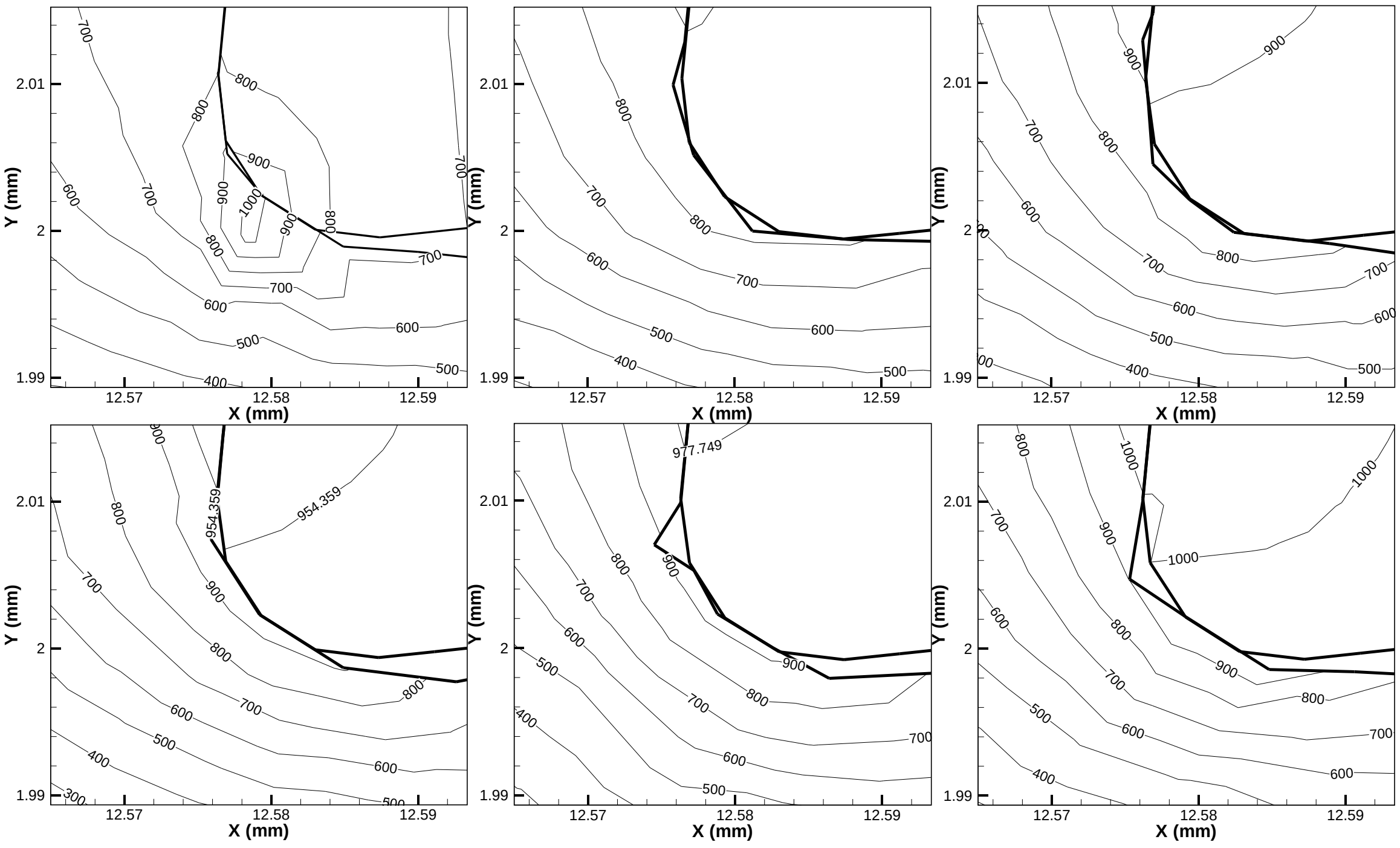

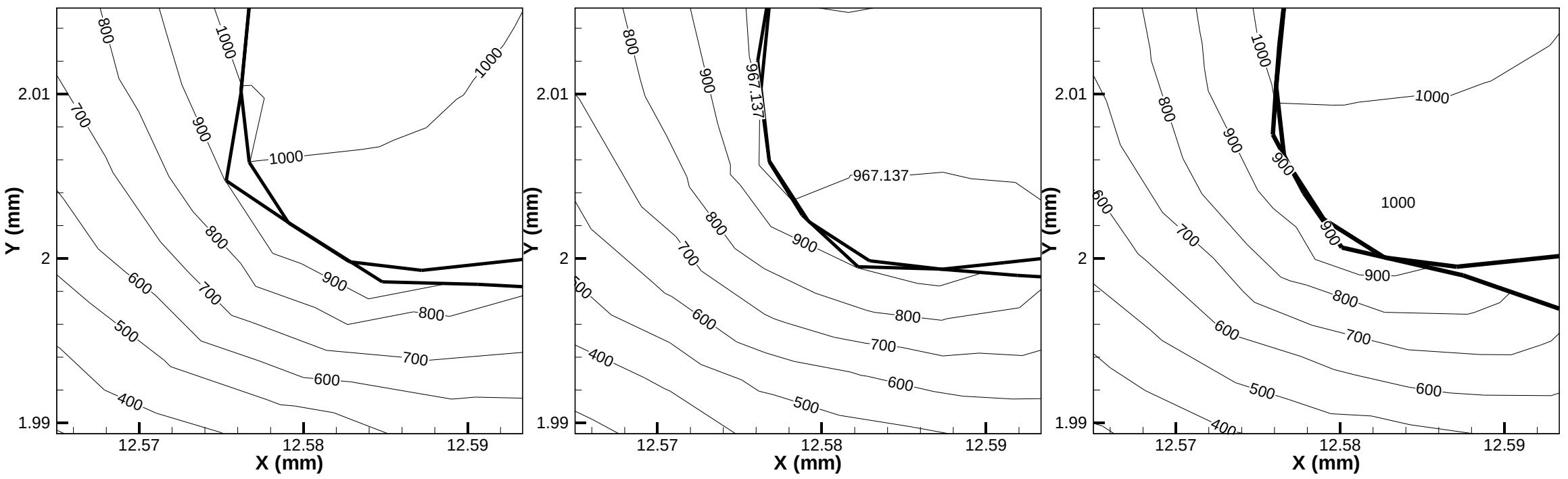

Figure 4: Temperature variation with time of machining

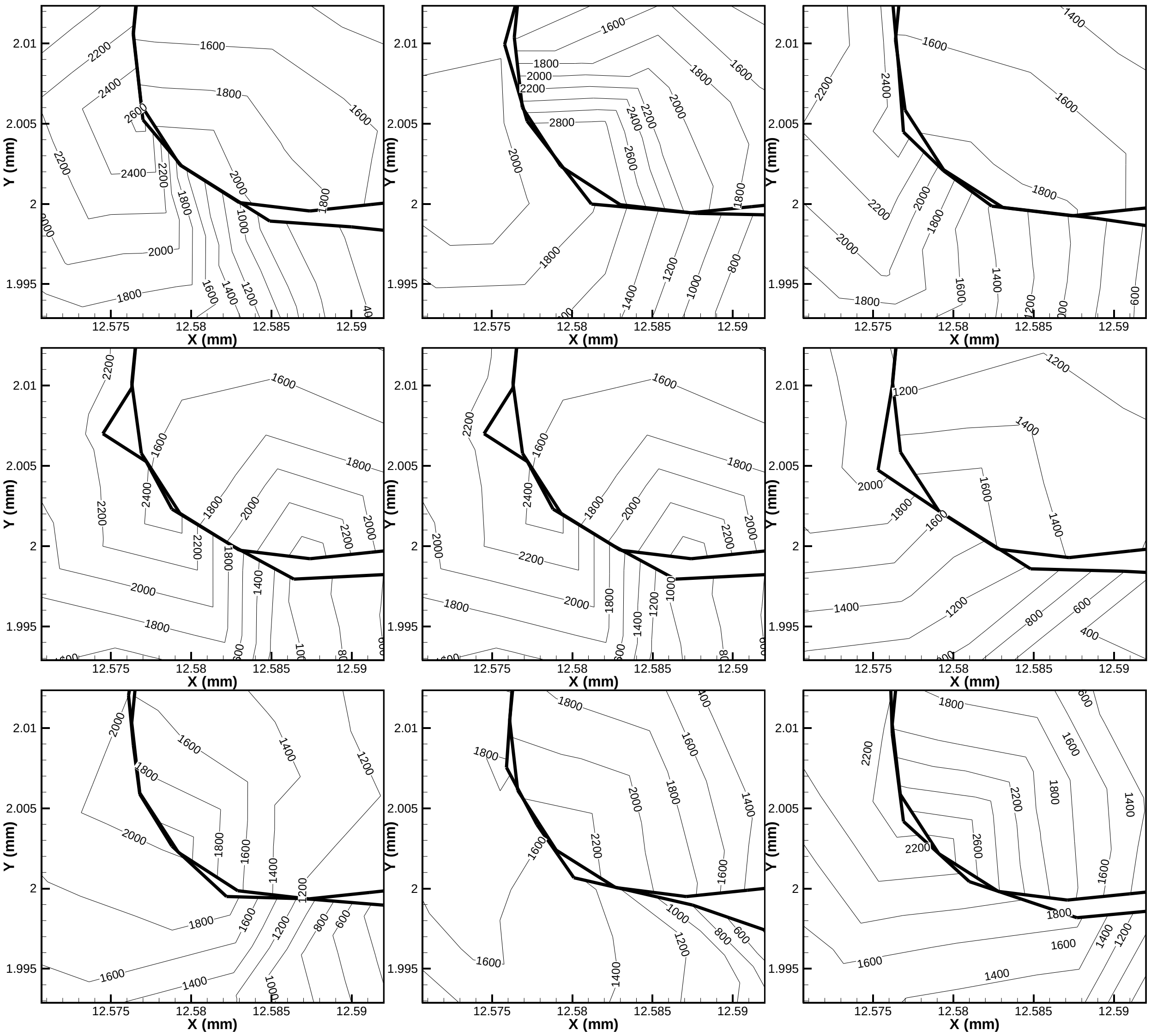

Figure 5 pressure evolution

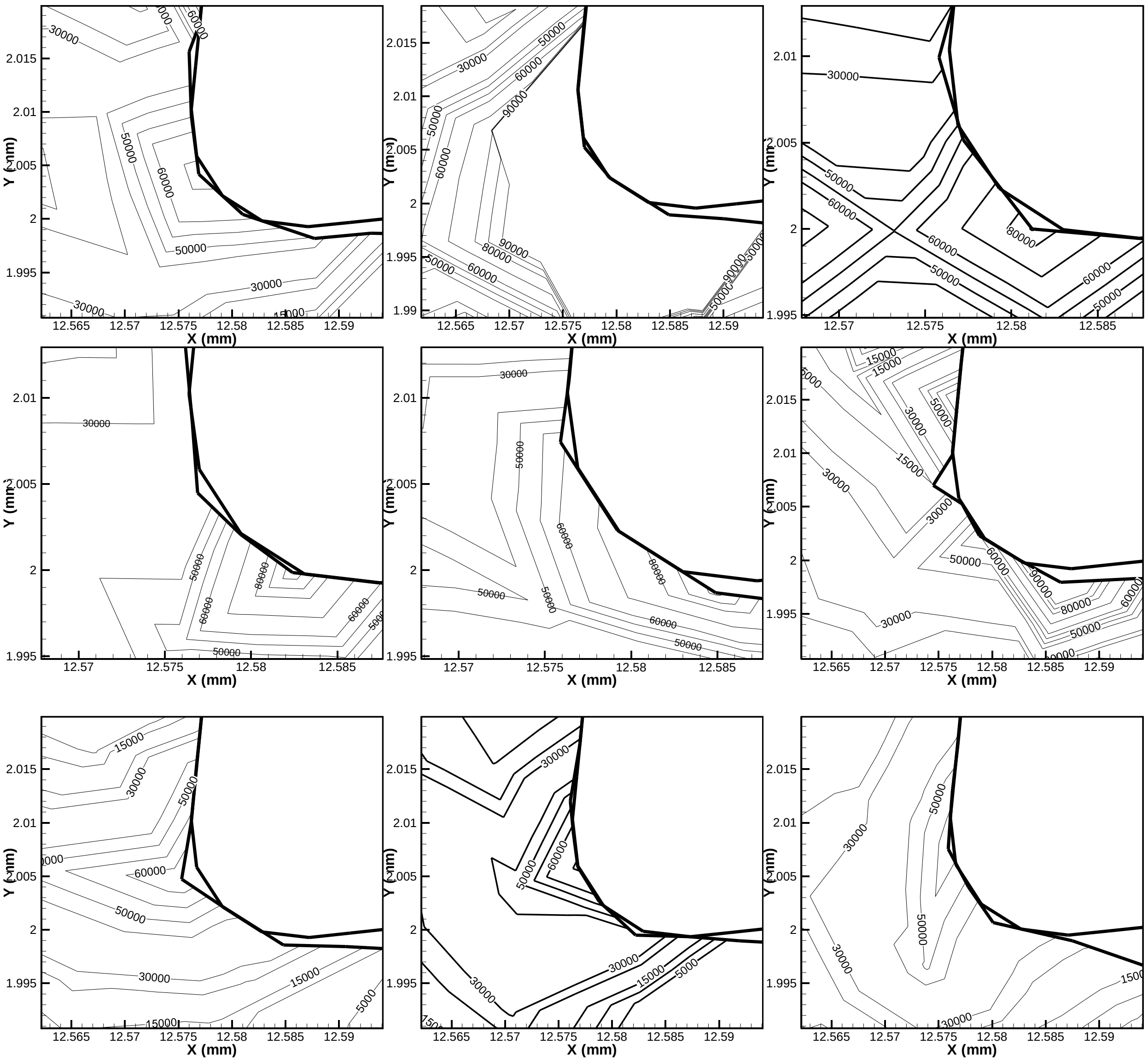

Figure 7: Evolution of the rate of heat release

## Evolution of Transport Properties in The Active Zone

To probe the evolution of thermal conduction in response to evolution of pressure within the active zone, three points on the tool-workpiece interface are studied. The location of the points is chosen such that they envelope the tool-workpiece interface. Figure 8 illustrates the position of these points with respect to the tool contour, figure 8-a, and with respect to the tool-workpiece interface, figure 8-b. The first point, point A in figure 8, is located at the root the rake face of the tool toward the end of the curve describing tool nose. Point B, the second point, is located at the middle of the contact contour describing the interface between the tool nose and the workpiece. The third point, point C, is located at the root of the flank face of the tool where the workpiece embarks on losing contact with the tool. For each of these points data pertaining to pressure and temperature evolution, during the chip generation cycle, were extracted. The data was subsequently fed into equation three and the WFL to deduce the evolution of thermal conductivity and the electrical resistivity in the workpiece active zone.

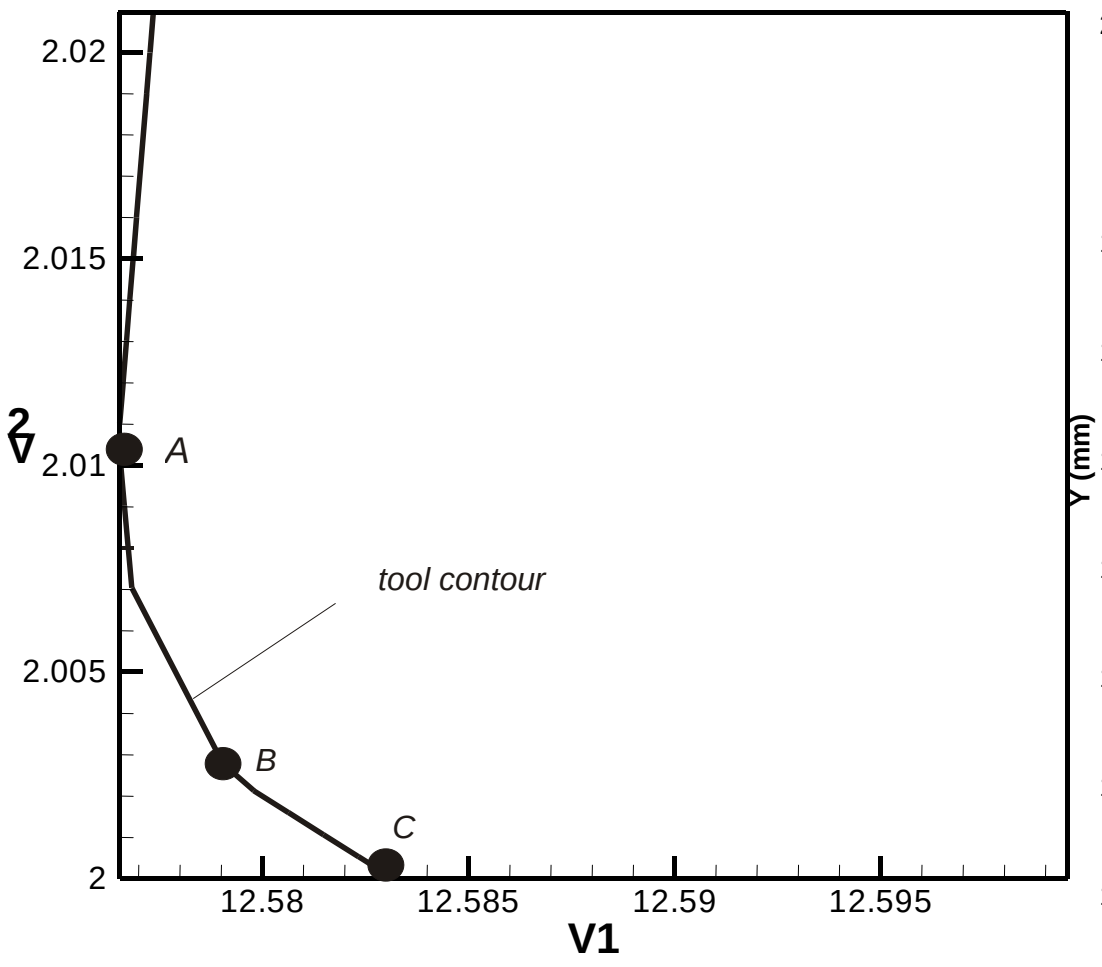

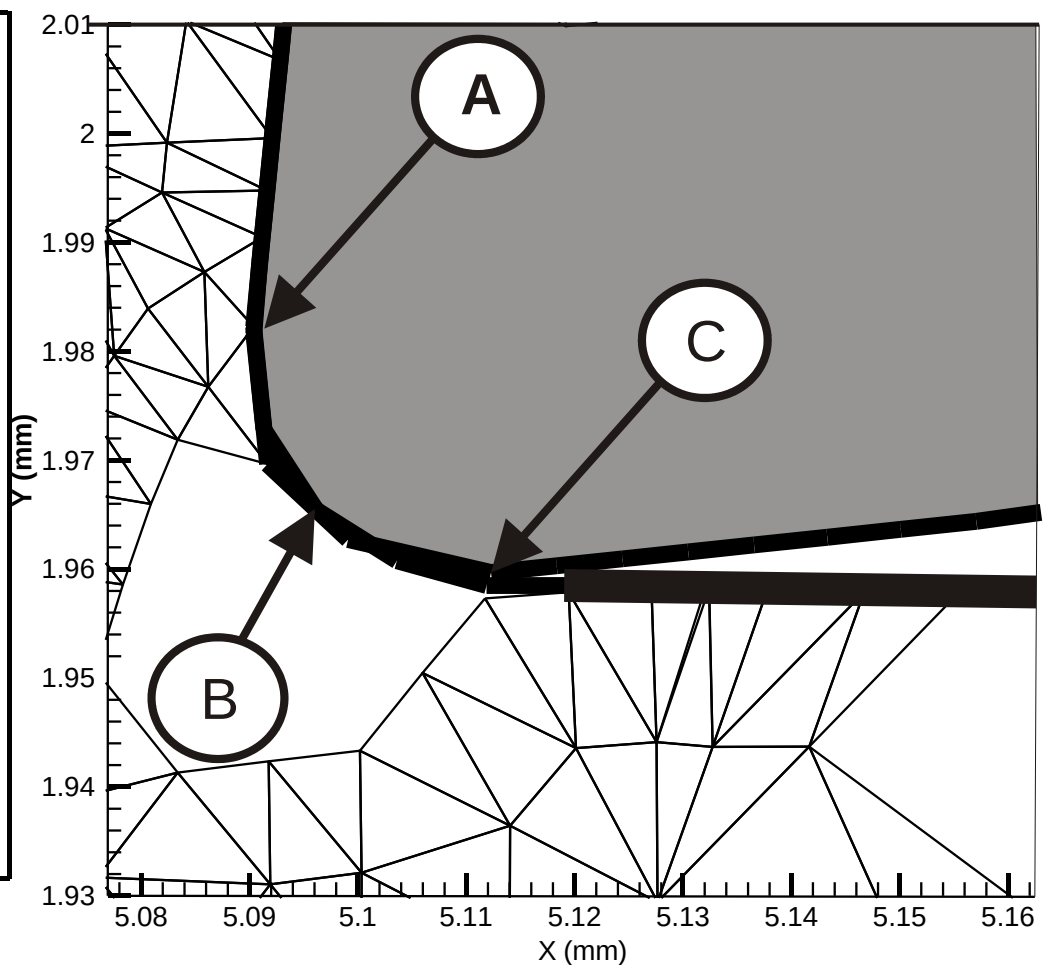

Figure 9-a depicts the evolution of temperature rise for each of the selected points plotted against the non-dimensional time within the chip generation cycle CGC. The temperatures generally increase with time. The rate of the increase, however, differs according to the phase in the CGC. Thus, three zones are identified. The first (analogous to the conception phase) is marked by an intense rate of climb. The temperatures within the incubation phase display a gradual more uniform increase rate. Finally, within the release phase the temperatures reach a plateau. That is the temperatures reach a quasi steady state.

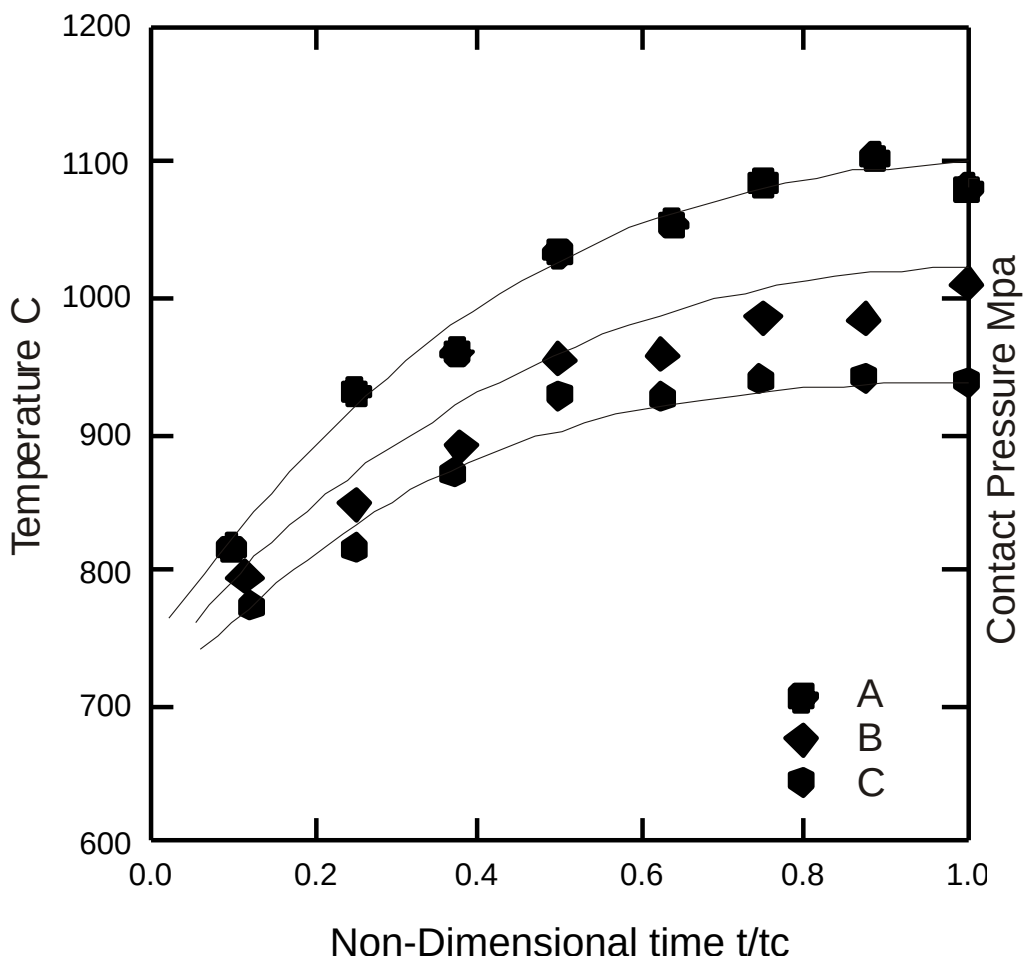

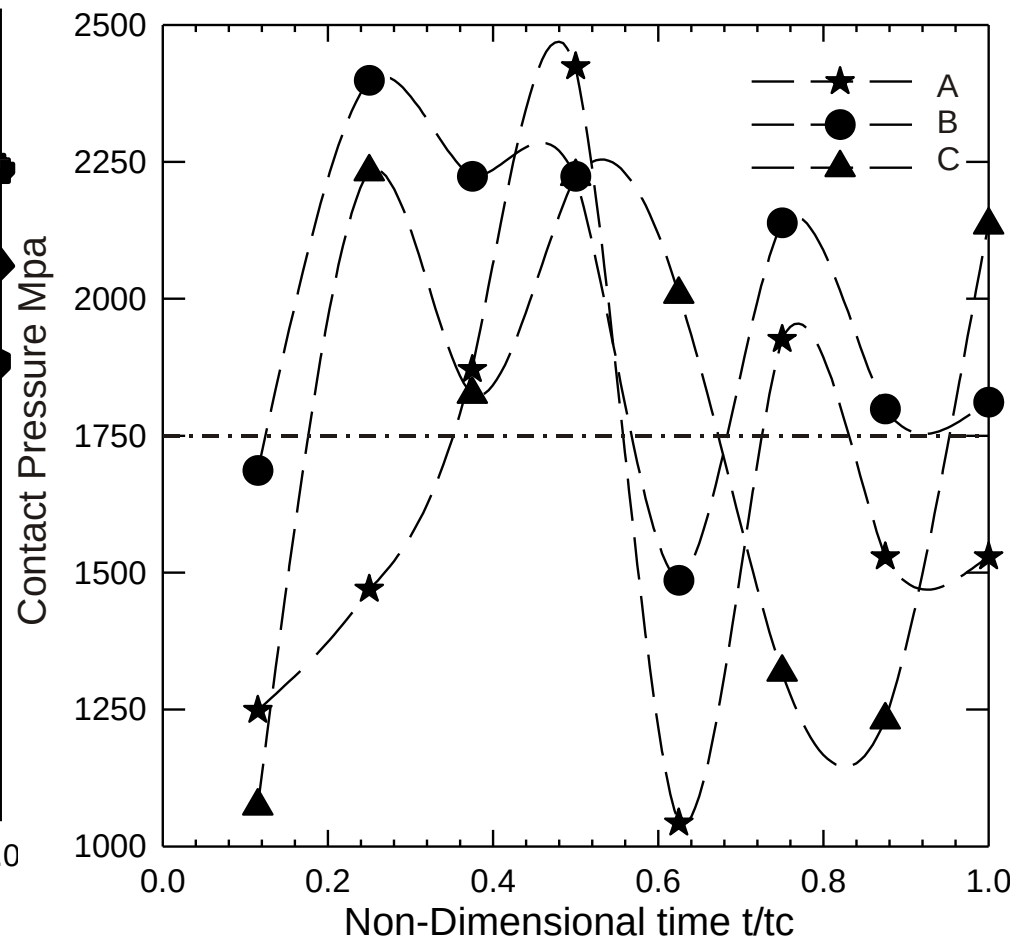

The evolution of pressure for all points is shown in figure 10. In contrast to the temperature evolution, the pressure displays a periodic trend. During conception of the chip the pressure rate of climb is rather steep especially for the rake face root point A (star symbol). In the following phase, incubation, the pressure drops in steep rate for the rake face point (albeit relatively less steep for the other points). Finally within the release phase, a contrasting behaviour is noticed. The pressure drops for the rake and contact points, points A and B, whereas pressure starts to climb for the flank face point, point C.

## Thermal conductivity

Variation in the thermal conductivity for the selected points is depicted in figure 11. A ratio of the thermal conductivity is plotted against time within the CGS. The ratio of conductivity represents the ratio between the actual conductivity, that which is a function of temperature and pressure K(P, T) to the thermal conductivity at standard conditions (room temperature and atmospheric pressure). The actual conductivity was calculated by means of equation three, using the temperature and pressure values extracted for each point (presented in the previous section).

It is noted that all points operate with an actual conductivity that is a fraction of the nominal conductivity ($0.6 \leq K(P, T)/ Ko \leq 0.8$. The conductivity of rake face boundary point, A, at the initiation of machining is sharply reduce. The conductivity reaches a plateau during the incubation phase, then recovers. Yet, that recovery is partial and not to the nominal state. The tool-nose point, B, and the flank face point C portray a slight departure from that trend. In particular, during the conception phase, the conductivity initially drops then partially recovers.

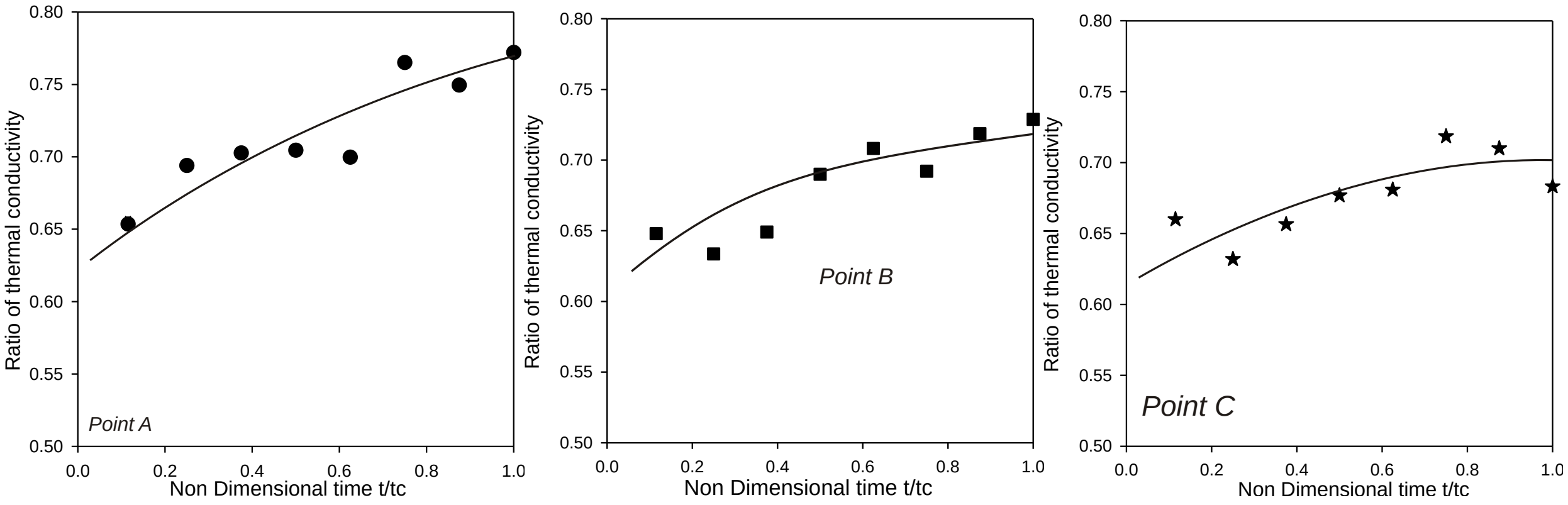

**Figure** : Evolution the Thermal Conductivity with pressure for the three representative points shown in figure (6).

The WFL law relates three quantities: the absolute temperature, the electrical conductivity, and the thermal conductivity. Data available in the current work are: the thermal conductivity, obtained by applying equation (3) and the temperatures, obtained from FE-simulations. Feeding the available data in the WFL yields the electrical resistivity (which is the reciprocal of the electrical conductivity, $\sigma$.

Values of the electrical resistivity resulting from applying the WFL are plotted as figure (8). The figure represents the evolution of the electrical resistivity, for the three points of interest represented in figure (7), as a function of the non-dimensional time of the chip generation cycle. Data for resistivity are normalized by the respective room-temperature room pressure values. The results indicate that the electrical resistivity for CPT exhibits an increase ranging between $1.1 \leq \rho\ (p, T)/\ \rho\ (p_o, T_o) \leq 1.25$ depending on the location on the tool nose. Maximum value of such increase (15% - 25 %) is exhibited at point A, which is close to the rake face of the tool. Such a finding is interesting on the count that intense adhesion between workpiece material and tool material normally takes place on the flank face of the tool. An example of such occurrence is shown as figure (9).

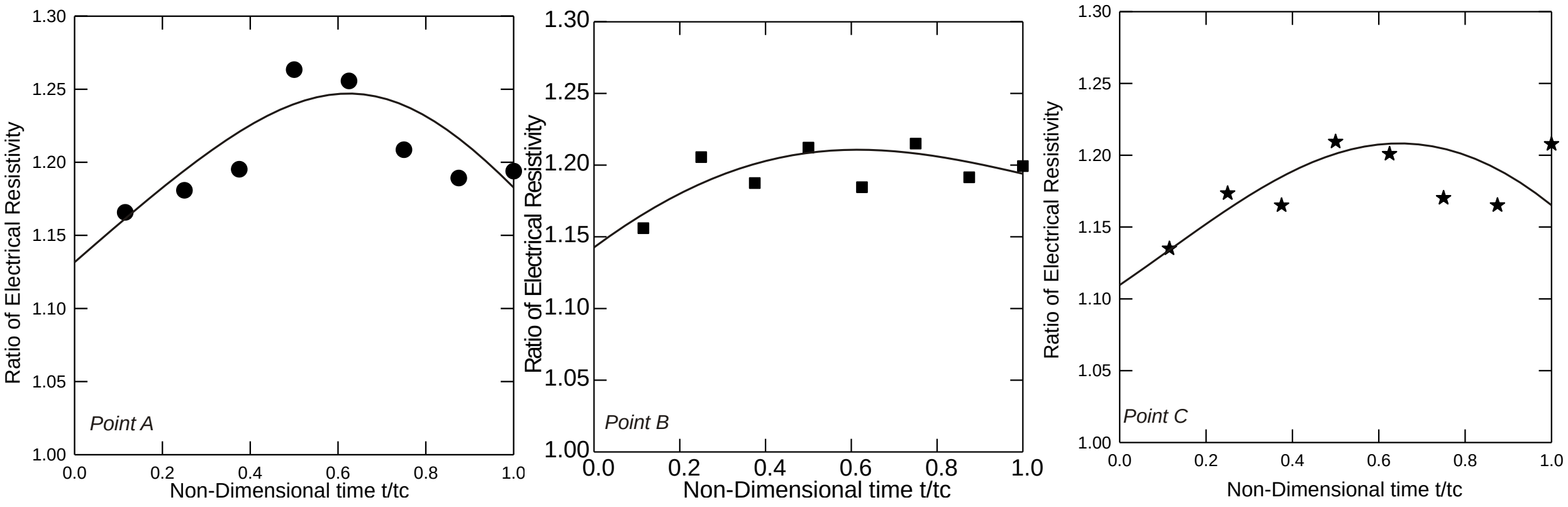

**Figure** : Electrical Resistivity

## Implications for Adhesion between Workpiece and Tool Material

The sliding of the created chip on the rake face of the tool may cause the generation of an electric current due to the phenomenon of "tribo-electrification". Akin to this current, an electrical potential will be created on both the tool and the workpiece. The strength of the resulting potential is may be calculated from [55],

$$\Phi(T)=\left[\frac{1}{2}\int_{T}^{T_{\max}}\rho(T)k(T)\,dT\right]^{1/2}$$

Equation (5) implies that the strength of the potential resulting from the sliding of the chip on the rake face depends on the current values of the electrical resistivity and thermal conductivity. These, in turn, are affected by the state of pressure.

It could be conjectured that material transfer from workpiece to tool is related to the strength of the resulting electrical potential and also on the gradient of that potential. That is it depends on whether the potential formed on the tool rake face is stronger than that on the sliding chip or vice versa. Further it may be conjectured that in general the material transfer will follow the direction of the potential gradient.

To investigate the feasibility of such a proposal, it is necessary to first evaluate the evolution of the strength of the potential at all the points of interest on the tool. Thus, if the strength of the electrical potential displays variation with time, it can be assumed that this variation may trigger material transfer to or from the workpiece.

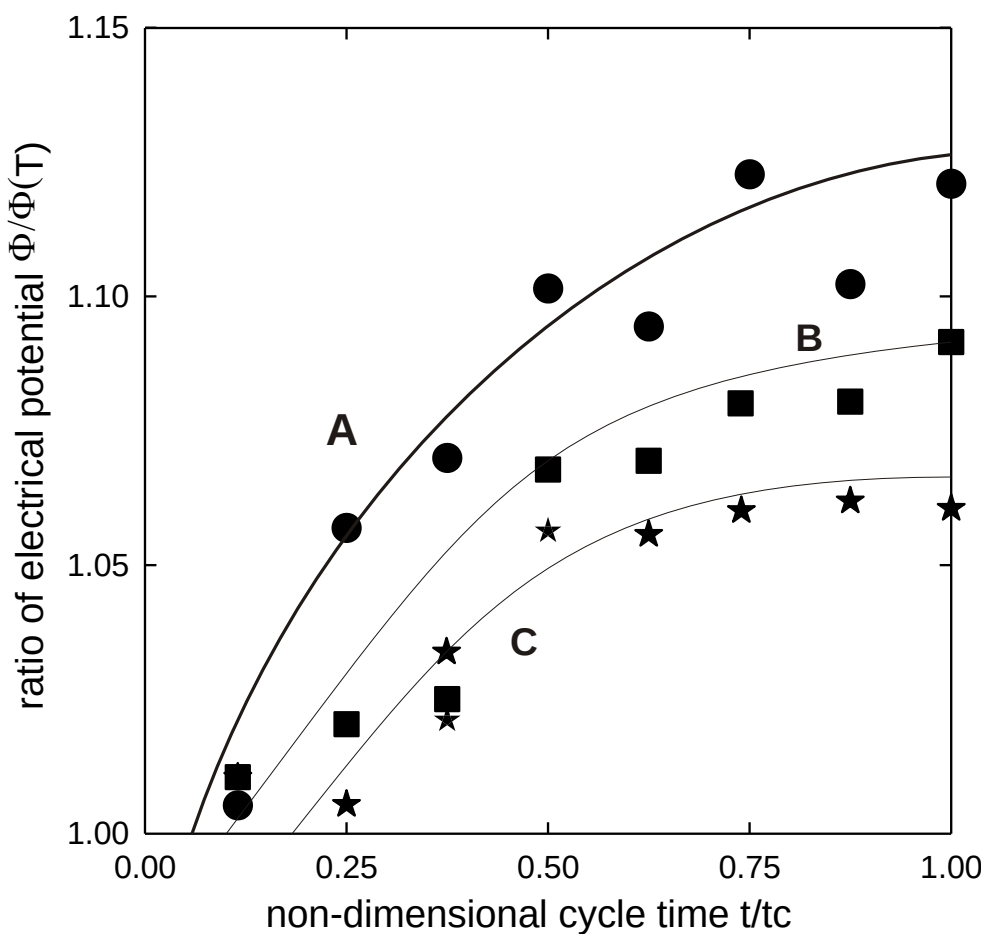

## DISCUSSION

The main objective of this work is to investigate the important role assumed by pressure-temperature combinations on the transport properties of the work piece while machining. The synergetic effect of such a combination was seen to dominant within an enclave of material directly located under the tool. This zone experiences an elevated stress state (pressure). It is also the incubator of the potentially generated surface and chips. Based on the derogatory influence of pressure on thermal conductivity of CPT one may conclude that behavior of the work piece, and the energetics of workpiece tool interaction will be governed by the value of pressure exerted upon tool engagement. Based on figure one, the initial contact pressure between tool and work piece is recognized as the primary factor that determines the trajectory of thermal and electrical transport in the active zone within the work piece. The freshly material engaging the tool will always be affected by the baric pressure history more than the thermal history of the contact. Such an effect is reported here for the first time in machining literature. It is never considered in previous numerical or theoretical studies of machining or tribology to the knowledge of the authors.

The importance of pressure-induced influence on thermal conduction drives from the coupling between thermal and electrical transport properties (especially in metals). A reduced ability to conduct heat (iie lower effective thermal conductivity) leads to higher resistance to electrical flow (at the same temperature). If one hypothesizes that the mode of damage to both tool and workpiece is related to the favorable mode of energy dissipation under the particular contact conditions, then the coupling between electrical and thermal transport becomes influential. When the ability of the active material zone to conduct heat is reduced, energy dissipation through electrical influences will be more active. That is, due to a reduced ability of thermal dissipation, electrical currents, and evolution of accompanying electro-static potentials, is more likely to be energetically favorable (or at least will be a viable option). If the strength of the potentials evolving on the tool side are different from those evolving on the work piece side, then a discharge force may take place in the direction of the potential gradient. Such a force may trigger stiction, adhesion or material transfer between tool and work piece. Details of such a conjecture, however, need more detailed studies.

Results obtained in this work indicate that damage, material transfer and thermal degradation mechanisms establish their level of activity in the first few moments of engagement between tool and work piece. Figures 7 through 9 indiacte that the intensity of thermal degradation, due to thermal or pressure effects, exhibits a steep increase during the chip conception phase. Whereas, the degradation increase uniformly until it reaches a plateaux during subsequent phases. The increase in electrical resistivity, and their by the increase in the strength of ESD potential,meanwhile exhibits and intense increase within the same phases. Interestingly, moreover, the length of adhesion between chip and workpiece, establishes its extent by the end of the conception phase. This implies that the most sustained damage, along with the intensity of damage inducing influences, are at their peak within the few initial moments of surface generation. That is, they are active intensely in the first moments of contact between tool and workpiece. The results also imply that the length of adhesion is a function of electrical activity on the tool and work piece surfaces. A connection between adhesion length and transport properties was hypothesized in the work of Friedman and lenz [ ] who reasoned that since the temperature affects the contact Length between chip and tool; and since the temperature itself is affected by thermal conduct levity, then thermal conductivity should affect the contact length. They further noticed that the location of the maximum temperature, in the contact length between chip and rake face, occurs in the trailing half of the contact. Such an area in their experiments represented the general area where transition from secondary deformation to sliding took place. The point at which the maximum temperature occurs, in our simulations is the transition point *A* (in figures 7 through 9). Such a point is characterized by a reduced thermal conductivity due to pressure induced effects (within the conception phase). It also exhibits the highest Level of ESD (figure 9). This suggests that electrical resistivity effects influence the contact length. In fact, most of the heat resulting from work piece deformation within-the secondary zone will accumulate withint the general area containing the transition points. Due to the drop of thermal conductivity, the dissipation of energy in this zone will be favored through electrical activity. In addition, due to the difference in strength between the potentials developed on the tool side and that developed on the side of the work piece, a force acting to increase the adhesion of the chip to the rake face will take place. The resulting heat will contribute toward softening of the chip material, thus increasing the contact area between chip and tool. An increased area, in light of a fixed chip width, leads to increased adhesion length.

The findings of this study point at the importance of revisiting design concepts of cutting tools. The revised concepts should consider, not only, mechanical influences but also thermal and thermo electrical effects. The authors submit that it is necessary to attempt to streamline the

## APPENDIX-1

## The Wiedemann-Franz-Lorenz law

The electron fluid in a metal is an excellent conductor of electrical charge. It is also an excellent conductor of heat. In an insulator, phonons (quantized lattice vibration) carry all the heat current. By contrast, in familiar metals, the electron fluid conducts nearly the entire heat current (the phonon current still exists but constitutes an extremely small percentage of the total heat current). Although the electrical and thermal currents carry distinct quantities (charge and entropy) it is fruitful to compare their magnitudes. The earliest comparison of these two quantities, the thermal conductivity K and the electrical conductivity $\sigma$ was made by Wiedemann and Franz in 1853.

Wiedemann and Franz used a classical approach based upon the fact that the heat and electrical transport both involve the free electrons in the metal. The thermal conductivity increases with the average particle velocity since that increases the forward transport of energy. However, the electrical conductivity decreases with particle velocity increases because of the collisions divert the electrons from forward transport of charge. This means that the ratio of thermal to electrical conductivity depends upon the average velocity squared, which is proportional to the kinetic temperature. Based on their classical mechanics approach Wiedemann and Franz deduced that the ratio of thermal to electrical conductivity is given by:

$$\frac{K}{\sigma T} = 2.45*10^{-8} \quad W\Omega K^{-2}$$

The constant in equation (3) was later corrected by Lorentz who confirmed the original findings of Wiedemann and Franz, albeit using quantum mechanics approach. It is to be noted however, that the constant in equation (3) is universal and is not affected by pressure [99].